\documentclass[a4paper]{jpconf}
\usepackage{graphicx, natbib}
\newcommand{\apj}[2]{{\em Astrophys. J.}, {\bf #1}, #2}
\newcommand{\apjl}[2]{{\em Astrophys. J. Lett.}, {\bf #1}, #2}

\newcommand{\mnras}[2]{{\em Mon. Not. R. Astr. Soc.}, {\bf #1}, #2}
\newcommand{\prd}[2]{{\em Phys. Rev. D}, {\bf #1}, #2}

\begin{document}
\title{Pulsar Timing Arrays:  No longer a Blunt Instrument for Gravitational Wave Detection}

\author{Andrea N. Lommen}

\address{Franklin and Marshall College, 415 Harrisburg Pike, Lancaster, PA 17604}

\ead{andrea.lommen@fandm.edu}

\begin{abstract}
Pulsar timing now has a rich history in placing limits on the stochastic background of gravitational waves, and we plan
soon to reach the sensitivity where we can detect, not just place limits on, the stochastic background.  However, the capability
of pulsar timing goes beyond the detection of a background.  Herein I review efforts that include single source
detection, localization, waveform recovery, a clever use of a ``time-machine" effect, alternate theories of gravity, 
and finally studies of the noise in our ``detector" that will allow us to tune and optimize the experiment.  Pulsar
timing arrays are no longer ``blunt" instruments for gravitational-wave detection limited to only detecting an
amplitude of the background.  Rather they are shrewd and tunable detectors, capable of a rich and dynamic variety of
astrophysical measurements.

\end{abstract}

\section{Introduction}

Pulsars are basically celestial clocks, and as such, can be used to construct a Galactic-scale gravitational wave 
detector using the same concept as ground-based interferometric detectors, i.e. one looks for phase changes in
the arrival of the signal at the vertex station, in this case, earth.  The length scales of our detector `arms' 
(1000 light years) as compared to the length of ground-based arms (4km) allow us to probe a different 
gravitational-wave frequency regime (nHz), a complement to the ground-based kHz 
regime \citep{Yardley10}.  For almost 30 years pulsar timers have been putting limits
on the energy density of the stochastic background using pulsar timing \citep{Romani83, Stinebring90, Kaspi94, Lommenthesis, 
Jenet06, vanHaasteren11,
Yardley11}. They point out that at some moment in the future, we will detect rather than limit the stochastic
background.  This moment is predicted to be sometime within this decade \citep{Demorest09, Verbiest09}.

In the last 10 years the field of gravitational-wave detection using pulsars has matured, and we are now considering
much more than just the background of gravitational waves.  We are demonstrating that very precise work on specific
sources can be done, and that we need to `tune' this detector in order to maximize our sensitivity to these sources.  This
manuscript briefly reviews these efforts, and 
is organized as follows. In \S \ref{sec:overview} I give some more details about the concept and current thought behind
using pulsar timing arrays (PTAs) to detect gravitational waves.  In the subsequent sections I 
review the work that shows that PTAs can
be 
(\S \ref{sec:directional}) 
directional detectors,  (\S \ref{sec:waveform}) used to recover the
gravitational waveform, 
(\S \ref{sec:time}) used to recover information
about the source at some time in past,
(\S \ref{sec:distance}) used to measure luminosity distance to gravitational-wave sources,
(\S \ref{sec:altgravity}) 
used to test alternate theories of gravity, and
(\S \ref{sec:noise}) characterized as a formal `detector' using measurements of their noise.
Finally, in \S \ref{sec:summary} I summarize the ways in which
a PTA is no longer a `blunt' instrument for gravitational-wave detection, but rather a tunable, pointable, and
adjustable detector that can be used to gain very specific astrophysical information about the gravitational-wave
source being detected.

\section{An overview of the concept of gravitational-wave detection using pulsars.}
\label{sec:overview}

\begin{figure}[h]
\includegraphics[width=28pc]{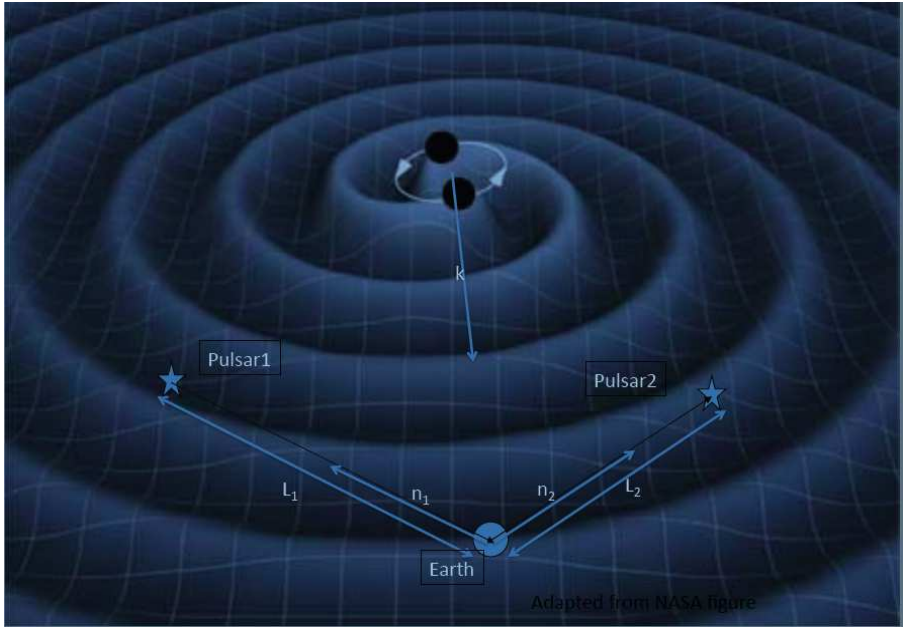}
\caption{\label{figure1}(Adapted from NASA) Schematic of a gravitational wave from a black hole binary impinging on a 2-pulsar pulsar timing array. 
When proper length scales are used the gravitational waves are nearly planar on the scale of the earth-pulsar systems, but \S \ref{sec:distance}
discusses the possibility of measuring their curvature.}
\end{figure}

Assume a gravitational wave is propagating through space in direction $\hat{k}$ apparently due to some distant source such as a
supermassive binary black hole (see figure 1).  
The gravitational wave changes the curvature of the space-time along which the electromagnetic wave is traveling, and as
such induces a change in the time that the pulse arrives at earth.  The size of the change at time $t$ for a pulse
from pulsar $j$, $\tau_{\rm GW}(\hat{k}, t)_j$, is given by
\begin{eqnarray}
\label{eq:tauj}
\tau_{GW}(\hat{k},t)_j &=
F^{+}(\hat{k},\hat{n}_j) g_{+}(t,L_j,\hat{k}\cdot\hat{n}_j) +
F^{\times}(\hat{k},\hat{n}_j) g_{\times}(t,L_j,\hat{k}\cdot\hat{n}_j),
\end{eqnarray}
for a TT-gauge gravitational-wave metric perturbation with form $h_{+}\mathbf{e}_{+}+h_{\times}\mathbf{e}_{\times}$.
$\hat{n}_j$ is a unit vector
pointing to pulsar $j$, $L_j$ is the distance to that pulsar.
$F^{+/\times}$ are geometric functions of $\hat{k}$ and $\hat{n}$ which we omit here for brevity but can be found
in \citet{Burt10}. 
Functions $g_{+}$ and $g_{\times}$ are integrals of $h_{+}$ and $h_{\times}$ as follows \citep{Finn10}:
\begin{eqnarray}
g_{(+/\times)}(t, L_j, \hat{k}_{j}\cdot \hat{n}^{j})=\int_0^{L_j} h_{+/\times} \left(t-(1+\hat{k}\cdot\hat{n}_j)(L_j-\lambda)\right)d\lambda.
\end{eqnarray}
Note that we are using geometrized units where $c=G=1$.
Following \citet{Finn10} we assume that a function $f$ exists for which
\begin{equation}
df_{+/\times}(u)/du = h_{+/\times}(u).
\end{equation}
For a plane wave we can then do the integral as follows:
\begin{eqnarray}
\label{earth_and_pulsar_terms}
g_{(+/\times)}(t, L_j, \hat{k}_{j}\cdot \hat{n}^{j})= \frac{f_{+/\times}(t)}{1 + \hat{k}\cdot\hat{n}_j} -  \frac{f_{+/\times}(t - (1 + \hat{k}\cdot \hat{n}_j)L_j)}{1 + \hat{k}\cdot\hat{n}_j}.
\end{eqnarray}
\noindent The first term is the so-called `earth term', and the second the `pulsar term'\citep{Jenet04}.
The pulsar term is delayed from the earth term by $(1 + \hat{k}\cdot \hat{n}_j)L_j$ which amounts to thousands of years in most cases.

Pulsar timers attempt to measure $\tau_{GW}$ by measuring the difference between
the expected arrival time of the pulse and the measured arrival time of the pulse.  This
difference is called a `residual.'  The residuals that we measure are certainly
not entirely due to passing gravitational waves, but to a variety of effects including
the interstellar medium \citep{vanStraten06, You07, Hemberger08}, measurement noise, and calibration errors \citep{Verbiest09}.
The gravitational-wave signature, however, has a distinct feature that none
of the other sources of noise can produce: part of the gravitational-wave signal, the earth
term, is correlated among all the pulsars.  
In other words, the earth term describes a response that all pulsar
timing residuals will exhibit at the same moment, independent of their distances.  
This characteristic can be used to distinguish
gravitational waves from other sources of residual \citep{Jenet05detect}.

The pulsar
term, however, is not correlated among the pulsars.  
For each pulsar, the
amplitudes of both terms will be modulated by the direction of the pulsar with respect
to the gravitational-wave source.  In addition, 
the temporal signature in the pulsar term is delayed by an 
amount which depends upon the distance to that pulsar.  The distance is generally not
known to better than 10\% which corresponds to hundreds of gravitational
wavelengths, and as such, it is basically a randomizer of the phase.  
One must confront this fact when attempting to recover the original
waveform and direction of the gravitational waves, as I discuss in the following
section.

However, before discussing PTAs as single source directional detectors, let us first consider
the combined effect of many sources.  This is actually the situation we expect
in the universe at large - hundreds of thousands of galaxy pairs merging, with their
central black holes eventually coming together to form black hole binaries.  Those black
hole binaries create a stochastic background of gravitational waves that we expect to
be able to detect in pulsar timing.

Though no stochastic background has been detected thus far, many authors have used pulsar
timing to limit the energy density of gravitational waves in the universe \citep{
Stinebring90, Kaspi94, Lommen01, Jenet05, vanHaasteren11}.  The most strict limit,
that placed by \citet{vanHaasteren11} places an upper limit on $A$ equal to
$6 \times 10^{-15}$ where $h_{\rm c} = A\left(\frac{f}{{\rm yr}^{-1}}\right)^{\alpha}$ and $\alpha=-2/3$.
which is at the boundary of existing models of galaxy merger rates \citep[see][and references therein]{Jaffe03, svc08}.
\cite{Wen11} use this to constrain the coalescence rate of supermassive black-hole binaries.
\nocite{vanHaasteren11} Van Haasteren et al. (2011) demonstrate that their analysis will eventually yield not just
the amplitude, but also the slope and perhaps even the shape of the spectrum.

Thus, limiting the stochastic background has been a necessary first step for the field, but there
is much to do beyond that.
PTAs can do much more than detect a single number such as the amplitude of the stochastic
background.  
They can be fashioned into a multi-faceted precision instrument for
gravitational-wave detection.

%

Armed with an understanding of the earth and pulsar terms, we can now proceed
to discuss the challenges of
the analysis using PTAs for gravitational-wave detection.
Basically, most methods of detection seek to 
capitalize on the coherence of the
earth term without being unduly hindered by the incoherence of the pulsar term.  
In the next section I review ways in which analyses of single sources go about
seeking the same advantage.

\section{PTAs as directional detectors}
\label{sec:directional}

As shown above, there are two terms in the response of a PTA:  the earth term and pulsar term.  
Given the location and waveform of a gravitational-wave source, the earth term in the equation above
is completely known.   We can turn that statement around and say that if we could somehow distinguish the earth term
separately from the pulsar term, and could then measure the amplitudes of the earth term in all our pulsars,
we could determine the location and waveform of the gravitational-wave source.  In most cases, we cannot separate
those two terms.  However, it turns out to be possible to determine the direction of the source using various
means described below.  The methods below all differ in the ways they have treated the pulsar term.

\citet{Finn10} looked at bursts of gravitational waves, sources whose duration is shorter than the
data span.  Pulsar timing data spans are tens of years, so a burst could be a month-long source.  
They put forth that by looking at bursts, one can ignore the pulsar term because it is
unlikely to enter the dataset in a human lifetime.  Note that because the distance between the earth
and the pulsar is hundreds or thousands of light years, the delay between the earth term and the pulsar
term is hundreds or thousands of years.  Thus, any source that produces a burst of coherent response in all the pulsars
(i.e. the earth term) will inflict its pulsar term on the same pulsars many hundreds of years later.
\citet{Finn10} were able to localize a strong source to less than 1 square degree.  
For a moderate source it was hundreds of square degrees. 
For a weak source it was thousands of square degrees. 

\nocite{Sesana10} Sesana and Vecchio (2010) looked at continuous gravitational-wave sources such as binary black holes, where one must include the pulsar term, and they
are able to localize the gravitational-wave source to within 40 square degrees for a 100-pulsar array and a signal to noise ratio (S/N) of 10.  
That number is big because they assumed they could not know the distances to the pulsars, so the 100\% variation of the pulsar term 
makes it very hard to pin down the direction as described at the beginning of this section.  
\cite{Ellis12} has achieved similar results working on a continuous wave pipeline.

\citet{Cornish10} claim they can actually search over and recover the pulsar distances from the chirp signal.  They have assumed 
Gaussian noise for the timing residuals which we may in fact have (see section on detector characterization below), 
but their technique has not been studied in the presence of red noise.
They are able to localize the source to less than 3 square degrees for strong sources.
\citet{Lee11single} point out 
that if the timing parallax (a distance measurement independent of the chirped signal) is estimated 
at the same time as the GW parameters are estimated
then the
pulsar term can be used as a great asset in increasing signal strength in single source cases.   
They carefully predict the statistical uncertainty that PTAs can expect to achieve in determining characteristics of
gravitational-wave sources such as orbital inclination angle, source position, frequency, and amplitude.
They predict that PTA source localization ability will range from a radian down to several microradians depending on the strength of the source.

Before I summarize these localization results, let me point out that
all the above work was done on simulated data in preparation for using real data.  However, work has been 
done on single gravitational-wave sources using real data.   
\citet{Yardley10} has used pulsar timing data to limit the number of coalescing binary systems
of a given chirp mass as a function of redshift.

My summary for single source localization is as follows.
We can localize well for strong burst sources, i.e.~less than 1 square degree.
For continuous wave sources it is more difficult because one cannot ignore the pulsar term.  
Without knowing the pulsar distances, it is not clear that one can do better than 40 square degrees, 
but there is hope for getting the pulsar distances either by mitigating the red noise in pulsars (see section
on detector characterization below) so that the distance can be found from the chirp signal \citep{Cornish10}, or by getting
the distances from other means such as timing or VLBI parallax \citep{Lee11single}.  
One interesting thing to note is that if one had an eccentric black-hole-binary instead of a circular one, it would be a repetitive burst source, 
and the localization would be much easier.

\section{PTAs as waveform recoverers}
\label{sec:waveform}
PTAs are potentially able to do much more than detect and localize a gravitational-wave source.  They may be able to actually recover the gravitational waveform.
This means that we would not just be able to pinpoint a source, but also actually decode the structure of the object that created it.  Things like masses,
spins, and orientations are encoded in these waveforms.  Direct observations would enable us to learn about the underlying structure of the black holes.

\citet{Finn10} demonstrate that one can recover the waveform of the gravitational-wave source provided the signal to noise ratio of the source
in the data is above about $1/10$ in each of about 30 pulsars.  They employ a likelihood method that does not rely on any
input template `bank.'   This could be used to recover astrophysical parameters of the source such
as inclination and polarization angles \citep{Sesana10}. 

\section{PTAs as time machines}
\label{sec:time}
\cite{Mingarelli11} show that 
when detecting a binary gravitational-wave source using pulsar timing, 
one can detect the source at two different moments in its orbit,  
because the earth and pulsar terms represent two different moments in its evolution.
If the black holes are spinning, for example, and there is spin-orbit precession in the system, the PTA
will measure two moments in its spin-orbit evolution at the same time.  
They can therefore probe higher order relativistic corrections in $(v/c)$, including the effect of spin-orbit coupling beyond the Newtonian approximation to the dynamics.

\section{PTAs as measurers of luminosity distance}
\label{sec:distance}
\cite{Deng11} demonstrate that the curvature of the waveform (seen in Figure 1) can be used to measure a gravitational-wave parallax, and thereby
obtain the luminosity distance to sources approaching or exceeding 100 Mpc.  This would serve as an important independent measurement of distance
to black hole binaries and other gravitational-wave sources.

\section{PTAs as detectors of alternate theories of Gravity}
\label{sec:altgravity}
In General Relativity, there are two transverse gravitational-wave polarization modes.  However, alternate theories of gravity can include
up to 6 gravitational-wave polarization modes.  
\citet{Lee08} find that if these extra polarizations exist, they could be detected in 5 years given 40-60 pulsars.
\citet{Chamberlin12} show that sensitivity to the vector
and scalar-longitudinal modes can increase dramatically for pulsar
pairs with small angular separations. For example, the
J1853+1303/J1857+0943 pulsar pair, with an angular separation of
about 3$^\circ$, is about 10,000 times more sensitive to a longitudinal
component of the stochastic background than to the transverse components.
Detecting an extra gravitational-wave polarization would provide the first evidence for the
violation of General Relativity.

Also in General Relativity, gravitational waves travel at the speed of light, and the graviton is therefore massless.  
However, in some alternate theories of gravity the graviton has mass, and the gravitational wave is therefore dispersive.
\citet{Lee10} estimate that it should be possible to detect this dispersion using PTAs.  In particular
they conclude that massless
gravitons can be distinguished from gravitons heavier than $3 \times 10^{-22}$ eV with 5 years of bi-weekly observations
of 60 pulsars with a pulsar rms timing accuracy of 100 ns.
This is not as good as limits placed currently by \citet{Goldhaber74} using galaxy cluster observations
 but the two methods are independent and pulsar timing would represent an important substantiation of the cluster results.

\section{Characterizing the Detector Noise}
\label{sec:noise}
As we make this transition to thinking about PTAs as shrewd and tunable detectors, one of the things we have had to confront is that
we must characterize the noise in the detector just as any other experiment would (See for example Cuocco et al. 2001)\nocite{Cuocco01}.   In our case the ``detector" is the collection of
pulsars. 
\citet{Shannon10}, \citet{Oslowski11}, \citet{Jenet11}, \citet{Perrodin12}, and \citet{Finn12} 
have all sought to characterize the noise in the pulsars.
\citet{Shannon10}, and \citet{Oslowski11} concentrate on determining what the ultimate sensitivity limit will be due to intrinsic variability
in the pulsars. \citet{Perrodin12} using the Cholesky transform as put forth by \citet{Coles11}
 concentrates on characterizing the current noise spectrum, while \citet{Finn12} uses a 
Bayesian analysis to identify the parameters of a noise model that best describe the timing noise statistics. 
\citet{Jenet11} give
a full error budget for pulsar timing considering various instrumental, propagation, and other fundamental sources
of noise.
The effort in all cases
is complicated by the fitting of the pulsar model to the time-of-arrival (TOA) data and by the fact that the pulsars are sampled at irregular intervals.  The
former renders stationary noise un-stationary, and the latter precludes the use of simple FFTs for finding the spectrum.  

Once the noise in the detector is characterized, we can optimize the detector.  \cite{Lee11} demonstrated the advantage of optimization which in our
case primarily consists of adjusting the amount of time spent on each pulsar, and showed that it only helps appreciably in the case of white noise.    Red noise
has been identified in many pulsar data sets \citep{Verbiest09, vanHaasteren11}, but recent techniques such as those by \citet{Demorest12} which
remove interstellar medium effects seem to significantly reduce the ``redness'' of the data.
In any case the possibility of ``tuning" the detector seems very promising.

\section{Summary}
\label{sec:summary}
PTAs have been long regarded as our hope for detecting gravitational waves in the nanoHertz band, but up until recently people have
thought that the best one could hope for would be a detection plus one or two numbers parameterizing the stochastic background in gravitational waves in
that regime (e.g., the spectral index and overall amplitude).  
We have started to be able to think about PTAs as detectors, with noise budgets, and various modes that we can tune
to suit the source in which we are most interested.  I have pointed to work here that is being done not only to detect gravitational
waves, but to use them for very particular goals including gravitational-waveform recovery, parameter estimation for
individual gravitational-wave sources, locating a source on the sky, 
distinguishing between theories of gravity, measuring luminosity distance to gravitational-wave sources,
and also using the
delayed pulsar term as something of a time machine, where we get a glimpse of the source as it existed in the past, and the present, all at once.

\ack
I would like to thank Sydney Chamberlin, Justin Ellis, Kejia Lee, Dick Manchester, Chiara Mingarelli, Willem van Straten, Daniel Yardley,
Joris Verbiest, and the entire detection group in NANOGrav, in particular Sydney Chamberlin, Sam
Finn, Xavi Siemens, and Delphine Perrodin for very useful discussions.


\begin{thebibliography}{}

\bibitem[{Burt}, {Lommen}, \& {Finn}(2011){Burt}, {Lommen}, and {Finn}]{Burt10}
{Burt}, B.~J., {Lommen}, A.~N., \& {Finn}, L.~S. 2011, \apj, 730, 17

\bibitem[Chamberlin \& Siemens(2012)Chamberlin and Siemens]{Chamberlin12}
Chamberlin, S., \& Siemens, X. 2012, submitted to ApJ

\bibitem[{Coles} {et~al.}(2011){Coles}, {Hobbs}, {Champion}, {Manchester}, and
  {Verbiest}]{Coles11}
{Coles}, W., {Hobbs}, G., {Champion}, D.~J., {Manchester}, R.~N., \&
  {Verbiest}, J.~P.~W. 2011, \mnras, 418, 561

\bibitem[{Corbin} \& {Cornish}(2010){Corbin} and {Cornish}]{Cornish10}
{Corbin}, V., \& {Cornish}, N.~J. 2010, ArXiv e-prints

\bibitem[{Cuoco} {et~al.}(2001){Cuoco}, {Losurdo}, {Calamai}, {Fabbroni},
  {Mazzoni}, {Stanga}, {Guidi}, and {Vetrano}]{Cuocco01}
{Cuoco}, E., {Losurdo}, G., {Calamai}, G., {Fabbroni}, L., {Mazzoni}, M.,
  {Stanga}, R., {Guidi}, G., \& {Vetrano}, F. 2001, \prd, 64, 122002

\bibitem[{Demorest} {et~al.}(2009){Demorest}, {Lazio}, {Lommen}, {Archibald},
  {Arzoumanian}, {Backer}, {Cordes}, {Demorest}, {Ferdman}, {Freire},
  {Gonzalez}, {Jenet}, {Kaspi}, {Kondratiev}, {Lazio}, {Lommen}, {Lorimer},
  {Lynch}, {McLaughlin}, {Nice}, {Ransom}, {Shannon}, {Siemens}, {Stairs},
  {Stinebring}, {Reitze}, {Shoemaker}, {Whitcomb}, and {Weiss}]{Demorest09}
{Demorest}, P., {Lazio}, J., {Lommen}, A., {Archibald}, A., {Arzoumanian}, Z.,
  {Backer}, D., {Cordes}, J., {Demorest}, P., {Ferdman}, R., {Freire}, P.,
  {Gonzalez}, M., {Jenet}, R., {Kaspi}, V., {Kondratiev}, V., {Lazio}, J.,
  {Lommen}, A., {Lorimer}, D., {Lynch}, R., {McLaughlin}, M., {Nice}, D.,
  {Ransom}, S., {Shannon}, R., {Siemens}, X., {Stairs}, I., {Stinebring}, D.,
  {Reitze}, D., {Shoemaker}, D., {Whitcomb}, S., \& {Weiss}, R. 2009, In
  astro2010: The Astronomy and Astrophysics Decadal Survey, ArXiv Astrophysics
  e-prints, p.~64

\bibitem[Demorest {et~al.}(2012)Demorest, Ferdman, Gonzalez, Nice, Ransom,
  Stairs, Arzoumanian, Cordes, Finn, Freire, Jenet, Kaspi, Lazio, Lommen,
  Lorimer, McLaughlin, Perrodin, Shannon, Siemens, and Stinebring]{Demorest12}
Demorest, P., Ferdman, R., Gonzalez, M., Nice, D., Ransom, S., Stairs, I.,
  Arzoumanian, Z., Cordes, J., Finn, L., Freire, P., Jenet, F., Kaspi, V.,
  Lazio, J., Lommen, A., Lorimer, D., McLaughlin, M., Perrodin, D., Shannon,
  R., Siemens, X., \& Stinebring, D. 2012, ApJ, in preparation

\bibitem[{Deng} \& {Finn}(2011){Deng} and {Finn}]{Deng11}
{Deng}, X., \& {Finn}, L.~S. 2011, \mnras, 414, 50

\bibitem[Ellis, Jenet, \& McLaughlin(2012)Ellis, Jenet, and
  McLaughlin]{Ellis12}
Ellis, J., Jenet, F., \& McLaughlin, M. 2012, in preparation

\bibitem[Finn(2012)Finn]{Finn12}
Finn, L.~S. 2012, in preparation

\bibitem[{Finn} \& {Lommen}(2010){Finn} and {Lommen}]{Finn10}
{Finn}, L.~S., \& {Lommen}, A.~N. 2010, \apj, 718, 1400

\bibitem[{Goldhaber} \& {Nieto}(1974){Goldhaber} and {Nieto}]{Goldhaber74}
{Goldhaber}, A.~S., \& {Nieto}, M.~M. 1974, \prd, 9, 1119

\bibitem[{Hemberger} \& {Stinebring}(2008){Hemberger} and
  {Stinebring}]{Hemberger08}
{Hemberger}, D.~A., \& {Stinebring}, D.~R. 2008, \apjl, 674, L37

\bibitem[{Jaffe} \& {Backer}(2003){Jaffe} and {Backer}]{Jaffe03}
{Jaffe}, A.~H., \& {Backer}, D.~C. 2003, \apj, 583, 616

\bibitem[{Jenet} {et~al.}(2004){Jenet}, {Lommen}, {Larson}, and {Wen}]{Jenet04}
{Jenet}, F.~A., {Lommen}, A., {Larson}, S.~L., \& {Wen}, L. 2004, \apj, 606,
  799

\bibitem[{Jenet} {et~al.}(2005){Jenet}, {Hobbs}, {Lee}, and
  {Manchester}]{Jenet05detect}
{Jenet}, F.~A., {Hobbs}, G.~B., {Lee}, K.~J., \& {Manchester}, R.~N. 2005,
  \apjl, 625, L123

\bibitem[{Jenet}, {Creighton}, \& {Lommen}(2005){Jenet}, {Creighton}, and
  {Lommen}]{Jenet05}
{Jenet}, F.~A., {Creighton}, T., \& {Lommen}, A. 2005, \apjl, 627, L125

\bibitem[{Jenet} {et~al.}(2006){Jenet}, {Hobbs}, {van Straten}, {Manchester},
  {Bailes}, {Verbiest}, {Edwards}, {Hotan}, {Sarkissian}, and {Ord}]{Jenet06}
{Jenet}, F.~A., {Hobbs}, G.~B., {van Straten}, W., {Manchester}, R.~N.,
  {Bailes}, M., {Verbiest}, J.~P.~W., {Edwards}, R.~T., {Hotan}, A.~W.,
  {Sarkissian}, J.~M., \& {Ord}, S.~M. 2006, \apj, 653, 1571

\bibitem[{Jenet}, {Armstrong}, \& {Tinto}(2011){Jenet}, {Armstrong}, and
  {Tinto}]{Jenet11}
{Jenet}, F.~A., {Armstrong}, J.~W., \& {Tinto}, M. 2011, \prd, 83, 081301

\bibitem[{Kaspi}, {Taylor}, \& {Ryba}(1994){Kaspi}, {Taylor}, and
  {Ryba}]{Kaspi94}
{Kaspi}, V.~M., {Taylor}, J.~H., \& {Ryba}, M.~F. 1994, \apj, 428, 713

\bibitem[{Lee} {et~al.}(2010){Lee}, {Jenet}, {Price}, {Wex}, and
  {Kramer}]{Lee10}
{Lee}, K., {Jenet}, F.~A., {Price}, R.~H., {Wex}, N., \& {Kramer}, M. 2010,
  \apj, 722, 1589

\bibitem[{Lee}, {Jenet}, \& {Price}(2008){Lee}, {Jenet}, and {Price}]{Lee08}
{Lee}, K.~J., {Jenet}, F.~A., \& {Price}, R.~H. 2008, \apj, 685, 1304

\bibitem[{Lee} {et~al.}(2011){Lee}, {Wex}, {Kramer}, {Stappers}, {Bassa},
  {Janssen}, {Karuppusamy}, and {Smits}]{Lee11single}
{Lee}, K.~J., {Wex}, N., {Kramer}, M., {Stappers}, B.~W., {Bassa}, C.~G.,
  {Janssen}, G.~H., {Karuppusamy}, R., \& {Smits}, R. 2011, \mnras, 414, 3251

\bibitem[Lee {et~al.}(2011)Lee, Bassa, Janssen, Karuppusamy, Kramer, Smits, ,
  and Stappers]{Lee11}
Lee, K.~J., Bassa, C.~G., Janssen, G.~H., Karuppusamy, R., Kramer, M., Smits,
  R., , \& Stappers, B.~W. 2011, \mnras, submitted

\bibitem[Lommen(2001)Lommen]{Lommenthesis}
Lommen, A. 2001, PhD Thesis, UC Berkeley

\bibitem[{Lommen} \& {Backer}(2001){Lommen} and {Backer}]{Lommen01}
{Lommen}, A.~N., \& {Backer}, D.~C. 2001, \apj, 562, 297

\bibitem[Mingarelli {et~al.}(2011)Mingarelli, Grover, Sidery, Smith, and
  Vecchio]{Mingarelli11}
Mingarelli, C. M.~F., Grover, K., Sidery, T., Smith, R. J.~E., \& Vecchio, A.
  2011, in preparation

\bibitem[{Os{\l}owski} {et~al.}(2011){Os{\l}owski}, {van Straten}, {Hobbs},
  {Bailes}, and {Demorest}]{Oslowski11}
{Os{\l}owski}, S., {van Straten}, W., {Hobbs}, G.~B., {Bailes}, M., \&
  {Demorest}, P. 2011, \mnras, 418, 1258

\bibitem[Perrodin {et~al.}(2012)Perrodin, Demorest, Finn, Jenet, Lommen, Siemens, and
  Romano]{Perrodin12}
Perrodin, D., Demorest, L.~S. Finn, P., Jenet, F., Lommen, A., Siemens, X., \& Romano, J.
  2012, in preparation

\bibitem[{Romani} \& {Taylor}(1983){Romani} and {Taylor}]{Romani83}
{Romani}, R.~W., \& {Taylor}, J.~H. 1983, \apjl, 265, L35

\bibitem[{Sesana} \& {Vecchio}(2010){Sesana} and {Vecchio}]{Sesana10}
{Sesana}, A., \& {Vecchio}, A. 2010, \prd, 81, 104008

\bibitem[{Sesana}, {Vecchio}, \& {Colacino}(2008){Sesana}, {Vecchio}, and
  {Colacino}]{svc08}
{Sesana}, A., {Vecchio}, A., \& {Colacino}, C.~N. 2008, \mnras, 390, 192--209

\bibitem[{Shannon} \& {Cordes}(2010){Shannon} and {Cordes}]{Shannon10}
{Shannon}, R.~M., \& {Cordes}, J.~M. 2010, \apj, 725, 1607

\bibitem[{Stinebring} {et~al.}(1990){Stinebring}, {Ryba}, {Taylor}, and
  {Romani}]{Stinebring90}
{Stinebring}, D.~R., {Ryba}, M.~F., {Taylor}, J.~H., \& {Romani}, R.~W. 1990,
  Physical Review Letters, 65, 285

\bibitem[{van Haasteren} {et~al.}(2011){van Haasteren}, {Levin}, {Janssen},
  {Lazaridis}, {Kramer}, {Stappers}, {Desvignes}, {Purver}, {Lyne}, {Ferdman},
  {Jessner}, {Cognard}, {Theureau}, {D'Amico}, {Possenti}, {Burgay},
  {Corongiu}, {Hessels}, {Smits}, and {Verbiest}]{vanHaasteren11}
{van Haasteren}, R., {Levin}, Y., {Janssen}, G.~H., {Lazaridis}, K., {Kramer},
  M., {Stappers}, B.~W., {Desvignes}, G., {Purver}, M.~B., {Lyne}, A.~G.,
  {Ferdman}, R.~D., {Jessner}, A., {Cognard}, I., {Theureau}, G., {D'Amico},
  N., {Possenti}, A., {Burgay}, M., {Corongiu}, A., {Hessels}, J.~W.~T.,
  {Smits}, R., \& {Verbiest}, J.~P.~W. 2011, \mnras, 414, 3117

\bibitem[{van Straten}(2006){van Straten}]{vanStraten06}
{van Straten}, W. 2006, \apj, 642, 1004

\bibitem[{Verbiest} {et~al.}(2009){Verbiest}, {Bailes}, {Coles}, {Hobbs}, {van
  Straten}, {Champion}, {Jenet}, {Manchester}, {Bhat}, {Sarkissian}, {Yardley},
  {Burke-Spolaor}, {Hotan}, and {You}]{Verbiest09}
{Verbiest}, J.~P.~W., {Bailes}, M., {Coles}, W.~A., {Hobbs}, G.~B., {van
  Straten}, W., {Champion}, D.~J., {Jenet}, F.~A., {Manchester}, R.~N., {Bhat},
  N.~D.~R., {Sarkissian}, J.~M., {Yardley}, D., {Burke-Spolaor}, S., {Hotan},
  A.~W., \& {You}, X.~P. 2009, \mnras, 400, 951

\bibitem[{Wen} {et~al.}(2011){Wen}, {Jenet}, {Yardley}, {Hobbs}, and
  {Manchester}]{Wen11}
{Wen}, Z.~L., {Jenet}, F.~A., {Yardley}, D., {Hobbs}, G.~B., \& {Manchester},
  R.~N. 2011, \apj, 730, 29

\bibitem[{Yardley} {et~al.}(2010){Yardley}, {Hobbs}, {Jenet}, {Verbiest},
  {Wen}, {Manchester}, {Coles}, {van Straten}, {Bailes}, {Bhat},
  {Burke-Spolaor}, {Champion}, {Hotan}, and {Sarkissian}]{Yardley10}
{Yardley}, D.~R.~B., {Hobbs}, G.~B., {Jenet}, F.~A., {Verbiest}, J.~P.~W.,
  {Wen}, Z.~L., {Manchester}, R.~N., {Coles}, W.~A., {van Straten}, W.,
  {Bailes}, M., {Bhat}, N.~D.~R., {Burke-Spolaor}, S., {Champion}, D.~J.,
  {Hotan}, A.~W., \& {Sarkissian}, J.~M. 2010, \mnras, 407, 669

\bibitem[{Yardley} {et~al.}(2011){Yardley}, {Coles}, {Hobbs}, {Verbiest},
  {Manchester}, {van Straten}, {Jenet}, {Bailes}, {Bhat}, {Burke-Spolaor},
  {Champion}, {Hotan}, {Oslowski}, {Reynolds}, and {Sarkissian}]{Yardley11}
{Yardley}, D.~R.~B., {Coles}, W.~A., {Hobbs}, G.~B., {Verbiest}, J.~P.~W.,
  {Manchester}, R.~N., {van Straten}, W., {Jenet}, F.~A., {Bailes}, M., {Bhat},
  N.~D.~R., {Burke-Spolaor}, S., {Champion}, D.~J., {Hotan}, A.~W., {Oslowski},
  S., {Reynolds}, J.~E., \& {Sarkissian}, J.~M. 2011, \mnras, 414, 1777

\bibitem[{You} {et~al.}(2007){You}, {Hobbs}, {Coles}, {Manchester}, {Edwards},
  {Bailes}, {Sarkissian}, {Verbiest}, {van Straten}, {Hotan}, {Ord}, {Jenet},
  {Bhat}, and {Teoh}]{You07}
{You}, X.~P., {Hobbs}, G., {Coles}, W.~A., {Manchester}, R.~N., {Edwards}, R.,
  {Bailes}, M., {Sarkissian}, J., {Verbiest}, J.~P.~W., {van Straten}, W.,
  {Hotan}, A., {Ord}, S., {Jenet}, F., {Bhat}, N.~D.~R., \& {Teoh}, A. 2007,
  MNRAS, 378, 493

\end{thebibliography}
\end{document}